\documentclass[12pt,preprint]{article}
\usepackage{epsf}
\newcommand{\eqb}{\begin{equation}}
\newcommand{\eqe}{\end{equation}}
\newcommand{\dmb}{\begin{displaymath}}
\newcommand{\dme}{\end{displaymath}}

\newcommand{\eab}{\begin{eqnarray}}
\newcommand{\eae}{\end{eqnarray}}
\newcommand{\ra}{\right\rangle}
\newcommand{\la}{\left\langle}

\newcommand{\be}{\begin{equation}}
\newcommand{\ee}{\end{equation}}

\begin{document}
\begin{titlepage}
\begin{flushright}
MPI-PhT 2001-34\\
\end{flushright}
\vspace{0.6cm}

\begin{center}
\Large{{\bf Non-perturbative non-locality: $V\pm A$ 
correlators in euclidean position space}}

\vspace{1cm}

Ralf Hofmann

\end{center}
\vspace{0.3cm}

\begin{center}
{\em Max-Planck-Institut f\"ur Physik\\ 
Werner-Heisenberg-Institut\\ 
F\"ohringer Ring 6, 80805 M\"unchen\\ 
Germany}
\end{center}
\vspace{0.5cm}

\begin{abstract}

Using an OPE modification, 
which takes non-perturbative non-locality into account, 
the difference and sum of vector and 
axial-vector correlators are evaluated in euclidean 
position space. For distances up to 0.8 fm 
the calculated behavior is close to the 
instanton liquid result and the ALEPH data, 
in contrast to the implications of the conventional OPE.

\end{abstract} 

\end{titlepage}

The determination of the momentum or distance range in which a truncated 
operator product expansion (OPE) in QCD approximates well 
is very important for many applications. 
Based on global quark-hadron duality 
the low-energy precision measurements of the spectral functions of vector 
and axial-vector channels (ALEPH experiment 
$\tau\rightarrow \nu_\tau+$hadrons \cite{Aleph}) could be confronted with 
the associated OPE's \cite{SVZ} but also with the instanton liquid model \cite{SS} 
and lattice calculations \cite{deGrand}.

In this letter we investigate how OPE-based 
non-perturbative non-locality affects the behavior of the $V\pm A$ 
correlators in euclidean position space. The results are compared 
to the instanton liquid data of \cite{SS} 
and the conventional OPE approach. 

Working in the chiral limit, in which both $\Pi^{V,A}_{\mu\nu}$ are transverse, and 
applying naive vacuum saturation at dimension 6, we obtain the 
following truncated OPE's \cite{SS} 
at a large normalization scale $Q_0$ 
\eab
\label{OPE}
\left.R_{V-A}(x)\right|_{Q_0}&\equiv&\left.\frac{\Pi^{V}(x)-\Pi^{A}(x)}{2\Pi_0(x)}\right|_{Q_0}=
\frac{\pi^3}{9}\,\alpha_s(Q_0)\la\bar{q}q\ra_{Q_0}^2\,\log[(xQ_0)^2]\,x^6\, ;\nonumber\\ 
\left.R_{V+A}(x)\right|_{Q_0}&\equiv&\left.\frac{\Pi^{V}(x)+\Pi^{A}(x)}{2\Pi_0(x)}\right|_{Q_0}=
1-\frac{\pi^2}{96}\la\frac{\alpha_s}{\pi}(F_{\mu\nu}^a)^2\ra_{Q_0}x^4
-\nonumber\\ 
& &\frac{2\pi^3}{81}\,\alpha_s(Q_0)\la\bar{q}q\ra_{Q_0}^2\,\log[(xQ_0)^2]\,x^6\ .
\eae
Thereby, we have only taken into account 
the leading-order $\alpha_s$ corrections in 
each mass dimension. $\Pi_0$ denotes the parton-model 
result which is equal 
for $V\pm A$. The $V-A$ correlator is free of 
purely perturbative contributions and only sensitive to chiral symmetry breaking. 
On the other hand, the $V+A$ correlator 
includes pure perturbation theory and moreover 
should directly reflect 
non-perturbative gluonic dynamics. Since the running of $\alpha_s$ almost cancels the 
log-powers at dimension 6, which come from the anomalous dimensions 
of the contributing operators, we can consider $\alpha_s$ to be fixed at scale $Q_0$ \cite{SVZ}. 

A modification of the conventional OPE was proposed in \cite{PRL} 
and applied to light-quark channels in \cite{PRD}. 
Since the idea is explained at length 
in these two papers we can be short here. For dimension 4 it was argued in 
\cite{PRL} that a finite correlation in the non-perturbative piece of 
the associated gauge invariant correlation function must 
decrease the relevance of fundamental 
field operators at low resolution. Coarse graining the vacuum expectation values (VEV's) of 
dimension 4 operators, an evolution equation was obtained 
which has the following solution in euclidean momentum space
\eqb
\label{run}
A(Q)=A(Q_0)\exp\left[-\frac{4}{5\lambda}\left(\frac{1}{Q}-\frac{1}{Q_0}\right)\right]\ .
\eqe
Thereby, $A(Q_0)$ is the VEV of a fundamental, 
gauge invariant composite and $Q_0$ the ``fundamentality'' 
scale down to which the 
description in terms of 
fundamental fields is sufficiently accurate. The (resolution independent) 
correlation length is denoted by $\lambda$. 

Assuming naive 
vacuum saturation \cite{SVZ} in the 
sense of {\bf 1)} in \cite{PRD}, 
the relevant 2 point functions 
to evaluate the modified version of eq.\,(\ref{OPE}) 
are the gauge 
invariant scalar quark and field strength correlators \cite{Dosch} 
which have been measured on the lattice \cite{DiGiacomo}. 
The following parameter values ($q$=quark, $g$=gluon) 
were obtained for $N_F=4$ (staggered) 
and a quark mass $a\cdot m=0.01$ 
\eab
\label{para}
\lambda_q&=&3.1\, \mbox{GeV}^{-1}\ ,\ \ \ \ A_q(a^{-1}=Q_0\sim 2\,\mbox{GeV})=(0.212\, \mbox{GeV})^3\ ;\nonumber\\ 
\lambda_g&=&1.7\, \mbox{GeV}^{-1}\ ,\ \ \ \ A_g(a^{-1}=Q_0\sim 2\,\mbox{GeV})=0.015\, (\mbox{GeV})^4\ .
\eae
\begin{figure}
\vspace{6.0cm}
\includegraphics{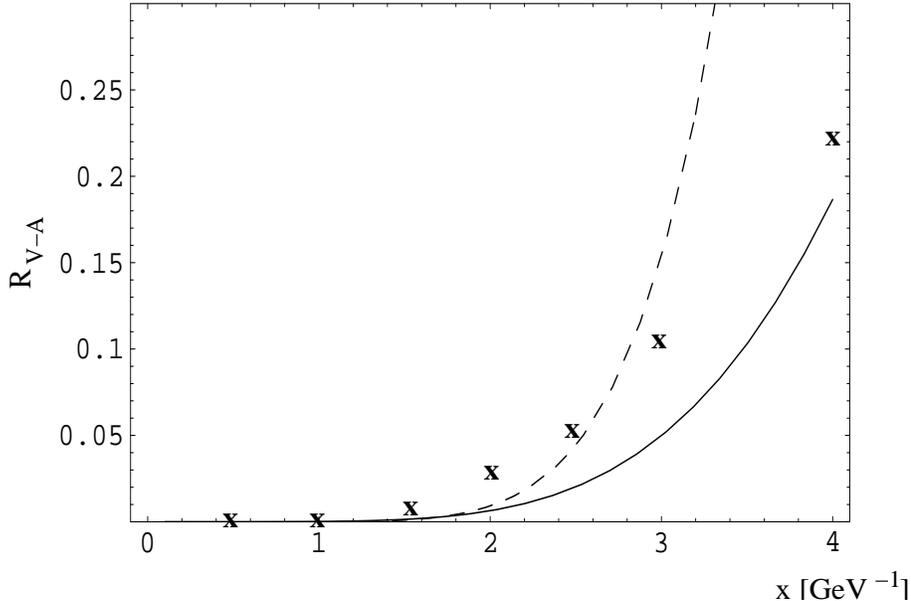}
\caption{$R_{V-A}$ as a function of distance $x$ in the chiral limit. 
The solid line corresponds to the non-perturbatively coarse 
grained OPE, the dashed line to the conventional OPE. Crosses depict the 
result of the instanton liquid 
calculation of \cite{SS} which is taken from \cite{deGrand}.} 
\label{} 
\end{figure}
Using (\ref{para}) and $\alpha_s(\mbox{2 GeV})=0.2$ \cite{Weisz} as 
inputs, appealing to eqs.\,(\ref{OPE},\ref{run}) (substituting $Q=1/x$ in (\ref{run})), 
we are now in a position to calculate $R_{V\pm A}$. 
Figs.\,1 and 2 show the results. For a comparison with the ALEPH data see ref.\,\cite{SS} where 
a spectral threshold $s_0$=2.5\,GeV$^2$ for 
the onset of perturbation theory is used in the $x$-space dispersion relation.    
\begin{figure}
\vspace{6.6cm}
\includegraphics{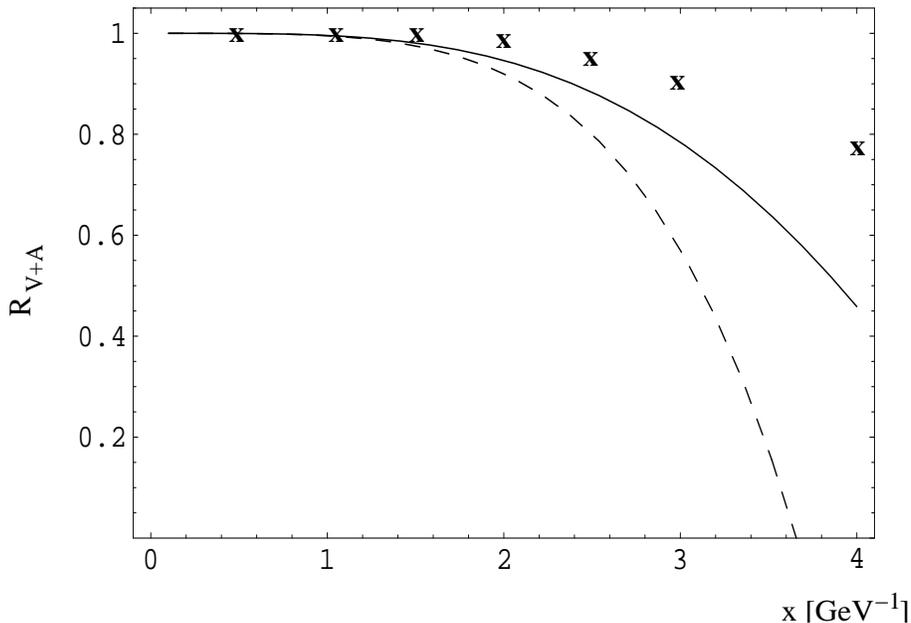}
\caption{Same as in Fig\,1., but now for $R_{V+A}$.} 
\label{} 
\end{figure}
Due to the moderation by exponential factors 
the power-like blow up of the non-perturbative corrections 
in conventional OPE's is not observed, and 
the modified OPE seems to converge also 
at larger distances than the 0.3 fm determined in \cite{SS}. 
Extrapolating the behavior of the lowest dimensions under non-perturbative 
coarse graining to higher dimensions $d<d_c$, 
each correction developes a maximum at $x_d=d\times\lambda_d$. 
Thereby, $d_c$ denotes the critical 
dimension of the asymptotic expansion 
which even the modified OPE is believed to be \cite{PRL}. 
It is a rather striking observation 
that our solid lines in Figs.\,1 
and 2 match the lattice results of ref.\,\cite{deGrand} 
almost perfectly. In quenched QCD this lattice calculation is based on the evaluation of the (truncated) overlap 
quark propagator which subsequently is used to build the correlators. 

To summarize we have shown 
that the concept of a non-perturbative component of 
OPE-based coarse graining leads in a wider range of distances 
to more realistic results for 
the euclidean $V\pm A$ correlators in position space 
than the conventional OPE's do.

\section*{Acknowledgements}    

The author would like to thank Uli Nierste for a stimulating conversation. Financial support 
from CERN's theory group during a research stay in June are gratefully acknowledged. 
The author is indebted to V. I. Zakharov for numerous useful 
discussions, valuable comments, and the encouragement to write this paper.

\bibliographystyle{prsty}

\begin{thebibliography}{10}



\bibitem{Aleph}
R. Barate et al.[ALEPH Collaboration], Z. Phys. {\bf C76}, 15 (1997);\\ 
Eur. Phys. J. {\bf C4}, 409 (1998).


\bibitem{SVZ}
M. Shifman, A. Vainshtein, and V. Zakharov, Nucl. Phys. B{\bf 147}, 385 (1979).

\bibitem{SS}
T. Sch\"afer and E. V. Shuryak, Phys. Rev. Lett. {\bf 86}, 3973 (2001).

\bibitem{deGrand}
T. deGrand, hep-lat/0106001.

\bibitem{PRL}
R. Hofmann, hep-ph/0109007.

\bibitem{PRD}
R. Hofmann, hep-ph/0109008.
     
\bibitem{Dosch}
H. G. Dosch, Phys. Lett. B {\bf 190}, 177 (1987).\\ 
H. G. Dosch and Yu. A. Simonov, Phys. Lett. B {\bf 205}, 339 (1988).

\bibitem{DiGiacomo}
M. D'Elia, A. Di Giacomo, and E. Meggiolaro, Phys. Lett. B{\bf 408}, 315 (1997).\\ 
A. Di Giacomo, E. Meggiolaro, and H. Panagopoulos, Nucl. Phys. B{\bf 483}, 371 (1997).\\ 
M. D'Elia,  A. Di Giacomo, E. Meggiolaro, Phys. Rev. {\bf D}{\bf 59}, (1999) 054503.

\bibitem{Weisz}
A. Bode et al., hep-lat/0105003.     


\end{thebibliography}

\end{document}